\documentclass[submission,copyright,creativecommons]{eptcs}
\providecommand{\event}{HAS 2013}

\usepackage{amsfonts}
\usepackage{amsmath}
\usepackage{amsthm}
\usepackage{color}
\usepackage{bm}
\usepackage{tikz} 
\usepackage{url} 
\usepackage{MnSymbol}
\usetikzlibrary{automata} 
\usepackage{algorithm,algorithmic}
\usepackage{breakurl}  

\input xy							
\xyoption{all}

\newtheorem{theorem}{Theorem}
\newtheorem{corollary}{Corollary}
\newtheorem{lemma}{Lemma}
\newtheorem{definition}{Definition}
\newtheorem{example}{Example}

\newcommand{\deq}{\stackrel{def}{=}}

\newcommand{\la}{\{\kern-1.5\nulldelimiterspace|}	
\newcommand{\ra}{|\kern-1.5\nulldelimiterspace\}}

\newcommand{\ls}{\la}	
\newcommand{\rs}{\ra}

\newcommand{\lsph}{(\kern-1.5\nulldelimiterspace|}	
\newcommand{\rsph}{|\kern-1.5\nulldelimiterspace)}

\newcommand{\lu}{\lfloor\!|}	
\newcommand{\ru}{|\!\rfloor }
\newcommand{\lo}{\lceil\! |}	
\newcommand{\ro}{|\!\rceil }

\newcommand{\lt}{\langle\kern-1.5\nulldelimiterspace|}	
\newcommand{\rt}{|\kern-1.5\nulldelimiterspace\rangle}

\newcommand{\lb}{\rsem}		
\newcommand{\rb}{\lsem}

\title{Approximated Symbolic Computations\\ over Hybrid Automata\thanks{
This work has been partially supported by Istituto Nazionale di Alta Matematica (INdAM).}}
\author{Alberto Casagrande
\institute{Dept. of Mathematics and Geosciences\\
University of Trieste, Italy}
\email{acasagrande@units.it}
\and
Tommaso Dreossi \qquad\qquad Carla Piazza
\institute{Dept. of Mathematics and Computer Science\\
University of Udine, Italy}
\email{\quad tommaso.dreossi@uniud.it\quad\qquad carla.piazza@uniud.it}
}
\def\titlerunning{Approximated Symbolic Computations over Hybrid Automata}
\def\authorrunning{A.~Casagrande, T.~Dreossi \& C.~Piazza}

\begin{document}

\maketitle
\begin{abstract}
	Hybrid automata are a natural framework for modeling and analyzing systems which exhibit a
	mixed discrete continuous behaviour.
However, the standard operational semantics defined over such models
implicitly assume perfect knowledge of the real systems and infinite precision measurements.
Such assumptions are not only unrealistic, but often lead to the construction of misleading models. For these reasons we believe
that it is necessary to introduce more flexible semantics able to manage with noise, partial information, and finite precision instruments. In particular, in this paper we integrate in a single framework based on approximated
	semantics different over and under-approximation techniques for hybrid automata.
	Our framework allows to both compare, mix, and generalize such techniques obtaining different
	approximated reachability algorithms.
\end{abstract}





\section{Introduction}
Hybrid automata were proposed to model hybrid systems, i.e., systems consisting of
interaction between discrete and continuous components \cite{Alur}. Automatic deduction of properties
for such systems is strictly related to the concept of state reachability. In particular, given a set of initial states,
we ask whether there are executions of the system that lead to specific final states. In general, it has been
proved that such problem is undecidable, i.e., algorithms which provide the correct answer for any instance 
of such problem cannot exist \cite{undecidable}. However, imposing syntactic restrictions, several subclasses of hybrid automata 
over which the reachability problem 
is decidable have been identified \cite{ominimal,focore2008}. 

Different approaches,
such as the introduction of \emph{noise in hybrid automata}~\cite{franzle99} and the use of approximated semantics
(\emph{$\epsilon$-semantics}~\cite{DEDS09}), have been proposed 
with the aim of both tackling the undecidability of the reachability problem and 
introducing hybrid automata semantics able of capturing some indeterminacy which is intrinsic in real world hybrid systems (e.g., experimental approximations, environmental disturbances, etc.).
In~\cite{DBLP:conf/hybrid/HenzingerR00}, the authors observed that undecidability cannot be removed by simply replacing trajectories with open flow tubes. More 
drastic changes at a semantics level are required. Many works proposed so far go in this direction (see \cite{franzle99,girard,tiwari,DEDS09}). 
Comparisons between these approaches on multistable and Zeno examples can be found in~\cite{hsb2012}. 

We proceed investigating in the direction of $\epsilon$, and more in general \emph{approximated}, semantics with the aim of introducing 
a general framework for the use, comparison, and composition of different approximation methods. 
Similarly to~\cite{franzle99}, our framework relies on polynomial dynamics. As shown by different authors (see e.g.~\cite{ourpaper,DBLP:conf/rtss/PrabhakarVVD09}),  
this is not a strong restriction since arbitrary flow functions can be approximated with polynomial flows. Moreover, the standard numeric algorithms 
used to generate the solutions of a system of differential equations and to integrate them are based on polynomials. 
Such considerations are exploited also in~\cite{ratschan09} where Fr\"anzle's results are applied ``\emph{[\ldots] after explicitly solving flow constraints [\ldots]}''.
While in~\cite{DBLP:conf/rtss/PrabhakarVVD09} the focus is on approximating hybrid systems with polynomial, here we start from polynomial systems 
and approximate their semantics with the aim of removing infinite precision and avoid unrealistic behaviors. 
Differently 
from~\cite{franzle99}  and~\cite{ratschan09}, we do not refer to robust systems. Indeed, multistable systems cannot 
be naturally modeled through robust automata~\cite{hsb2012}. 

In more details, in this work, we propose a framework that, on the one hand, 
exploits Fr\"anzle's approach, reformulated in terms of over-approximating semantics, 
and on the other, is based on $\epsilon$-semantics for under-approximating the reachable set.
Neither approach relies 
on fixed-grid discretizations, but on local perturbations on reachable points. 
In particular, in a quantum-physics fashion, $\epsilon$-semantics pertubates the observed states, but not 
the continuous evolution, which proceeds with infinite precision as long as it is not observed. 

We implemented our framework exploiting translations of approximated semantics into first-order formul\ae\ over the reals. Such translations allow us to exploit available tools for quantifier eliminations, such as \texttt{Redlog} and \texttt{QEPCAD}, to implement our reachability algorithms. We tested our implementation on a railroad crossing scenario. 

The paper is organized as follows. In Section~\ref{sec_ha} we present the notation and our definition of hybrid automaton. 
Section~\ref{sec_app} introduces the notion of approximating semantics and briefly reviews \emph{$\epsilon$-semantics}, while 
Section~\ref{sec_dist} instantiates \emph{disturbed automata} in our formalism.
In Section~\ref{mixing}, the standard semantics of disturbed automata and the related reachability algorithm are expressed in terms of an over-approximated semantics, while a $\epsilon$-semantics, under-approximating the standard one, is introduced. 
In Section~\ref{case}, we briefly describe our implementation of the presented framework and test it on the railroad case-study. 
Finally, Section~\ref{conclusions} ends the paper with some general comments on the difference between the compared approaches and 
suggests further developments. 


\section{Hybrid Automata}\label{sec_ha}

\subsection{Preliminaries}\label{sec_prel}
	We now introduce some notations and conventions. Capital letters $X,X',X_m$, and $X'_m$,
	where $m \in \mathbb{N}$, denote variables ranging over $\mathbb{R}$, while  
	$Z$ denotes the vector of variables $\langle X_1,\dots,X_d \rangle$ and $Z'$ denotes
	the vector $\langle X'_1,\dots,X'_d \rangle$. The variable $T$ 
	models time and ran\-ges over $\mathbb{R}_{\geq 0}$. We use $p,q,r,s,\dots$
	to denote $d$-dimensional vectors of real numbers.

	As far as the standard notions of first-order languages, models, and theories are concerned the reader may refer, for example, 
	to \cite{mendel}. In this paper we refer to the first-order theory of $(\mathbb R, 0,1,+,*,=,<)$, also known as the theory of 
    \emph{semi-algebraic sets} or Tarski's theory \cite{tarski}. Such theory is decidable, i.e., algorithms to check satisfiability of formul\ae\ have been defined (see, e.g., \cite{basu97}). 
    
We write $\varphi[X_1, \dots, X_m]$ to 
stress the fact that the set of free variables of the first-order formula 
$\varphi$ is included in the set of variables
$\{X_1$, $\ldots$, $X_m\}$.
If $\{Z_1$, $\ldots$, $Z_n \}$ is a set of variable 
vectors, $\varphi[Z_1$, $\ldots$, $Z_n]$ indicates that the free 
variables of $\varphi$ are included in the set of components of 
$Z_1$, $\ldots$, $Z_n$. Given a formula $\varphi[Z_1$, $\ldots$, $Z_i$, 
$\ldots$, $Z_n]$ 
and a vector $p$ of the same dimension as the variable vector $Z_i$, the 
formula obtained by component-wise substitution of $Z_i$ with $p$ is denoted 
by $\varphi[Z_1$, $\ldots$, $Z_{i-1}$,  $p$, $Z_{i+1}$, 
$\ldots$, $Z_{n}]$. 
   We use $\bot$ and $\top$ as shortcuts to denote the two formul\ae\ $0=1$ and $1=1$, respectively. 
    
The set of formal\ae\ having $n$ free-variables is denoted by $\mathcal F_n$, while 
$\ls \varphi \rs$, where $\varphi$ is any generic formula in $\mathcal F_n$, is 
the \emph{standard semantics} of $\varphi \in \mathcal F_n$ i.e.~the 
set of points of $\mathbb R^n$ satisfying $\varphi$. More formally, 
$
\ls \cdot \rs: \bigcup_{n\in\mathbb N} \mathcal F_n \rightarrow \bigcup_{n\in \mathbb N}\wp(\mathbb R^n)$
with
$\ls \varphi[X_1,\dots, X_n]\rs \deq \{\langle p_1,\dots,p_n\rangle\in \mathbb R^n \:|\: \varphi[p_1,\dots,p_n] \textrm{ holds}\}
$. 

On the other hand, given a set $\mathbb S\subseteq \mathbb R^n$ we say that a formula $S[Z]$ \emph{represents} (also \emph{defines}) $\mathbb S$ if
$\ls S[Z] \rs = \mathbb S$. Not all the subsets of $\mathbb R^n$ can be represented through a formula.


We also use some standard notions from topological and metric spaces (see~\cite{topology}).
Although we implicitly refer to the \emph{standard euclidean metric} $\delta$ over $\mathbb R^n$, 
our results can be generalized to any metric definable in Tarski's theory.
We write $B(p,\epsilon)$ to indicate the open sphere of radius $\epsilon$ centered in $p \in \mathbb R^n$. 
By extension, $B(\mathbb S,\epsilon)$, where $\mathbb S$ is a subset of $\mathbb R^n$, denotes  
the Minkowski sum of $B(0,\epsilon)$  and $\mathbb S$.
%

A set $\mathbb S$ is said to be \emph{$\alpha$-paraconvex}, where $\alpha\in [0,1]$, if for each $B(p,\epsilon)$ 
(with $p$ and $\epsilon$ generic) and for each $q\in conv(\mathbb S\cap B(p,\epsilon))$ it holds that $\delta(q,\mathbb S)\leq \alpha * \epsilon$, 
where $conv(\mathbb S\cap B(p,\epsilon))$ is the convex hull of 
$\mathbb S\cap B(\{p\},\epsilon)$ and $\delta(q,\mathbb S)=inf \{\delta(q,p) \:|\: p\in \mathbb S\}$.
%
Let $\mathcal{I} \subseteq \mathbb{R}$ be an interval and $f: \mathcal{I}\rightarrow \mathbb R^n$. We say that
$f$ is \emph{continuous} if for each $t\in \mathcal{I}$ and for each neighborhood $U_{f(t)}$ of $f(t)$ there exists
a neighborhood $U_t$ of $t$ in $\mathcal I$ such that for each $t'\in U_t$ it holds $f(t')\in U_{f(t)}$. 
		Moreover, a set-valued map $F : \mathcal{I} \rightarrow 
		\wp(\mathbb{R}^n)$ is \emph{lower semi-continuous} if for each $t \in \mathcal{I}$,
		for each $y \in F(t)$, and for each neighborhood $U_y$ of $y$, there exists a neighborhood 
		$U_t$ of $t$ in $\mathcal{I}$ such that for each $t' \in U_t$ it holds $F(t') \cap U_y \neq \emptyset$.
The notion of lower semi-continuity and $\alpha$-paraconvexity are at the basis of Michael's selection theorems (see, e.g., \cite{bookselection}) which guarantee the existence of a continuous flow inside a set-valued map. In particular, given a set-valued
map $F : \mathcal{I} \rightarrow \wp(\mathbb{R}^n)$ the \emph{selection problem} over $F$ requires to find (if there exists one) a continuous function 
$f:\mathcal{I} \rightarrow \mathbb{R}^n$ such that for each $t\in \mathcal I$ it holds that $f(t)\in F(t)$.

\subsection{Syntax}
In this section we give the formal definition of hybrid automata.
Many different definitions can be found in the literature. Most common differences between
those formalisms reside in the descriptions of continuous and discrete transitions,  while the semantics attributed to the transitions are almost the same. 
Here we follow the approach used in \cite{focore2008} and \cite{DEDS09} where automata are defined 
through first-order formul\ae\ over the reals and, in particular, semi-algebraic formul\ae.

\begin{definition}[Hybrid Automata - Syntax]\label{syn_ha}
		A \emph{hybrid automaton} $H$ $=$ $(Z,$ $Z',T,$ $\mathcal{V},$ $\mathcal{E},$ $Inv,$ $Dyn,$ $Act,$ $Res)$
		of dimension $d(H)$ consists of the following components:
		\begin{itemize}
			\item $Z=\langle X_1, \dots, X_{d(H)}\rangle$ and $Z' = \langle X'_1,\dots,X'_{d(H)} \rangle$
				are two vectors of variables ranging over the reals $\mathbb{R}$;
			\item $T$ is a variable ranging over $\mathbb{R}_{\geq 0}$;
			\item $\langle \mathcal{V}, \mathcal{E} \rangle$ is a finite directed graph. Each element of $\mathcal{V}$
				will be dubbed \emph{location};
			\item each vertex $v \in \mathcal{V}$ is labeled by the two formul\ae~$Inv(v)[Z]$
				and $Dyn(v)[Z,Z',T]$; it should holds that, when $Inv(v)[p]$ is true, 
				$Dyn(v)[p,q,0]$ is true if and only if $p = q$; 
			\item	 each edge $e \in \mathcal{E}$ is labeled by the two formul\ae~$Act(e)[Z]$ and
				$Res(e)[Z,Z']$.
		\end{itemize}
	\end{definition}

Intuitively, $Dyn(v)$ represents the dynamics associated to the location $v$, $Inv(v)$ denotes the 
set of continuous values admitted during the evolution in $v$,  $Act(e)$ identifies 
the set of continuous values from which the automaton can jump over the edge $e$, and 
$Res(e)$ characterizes a map that should be applied to the continuous values 
from which the automaton crosses the edge $e$. Section~\ref{sec:semantics} details 
the formal meaning of these formul\ae\ and describes the 
semantics of hybrid automata. 

While hybrid automaton dynamics are classically described by using differential equations 
(see, e.g., \cite{lafferiere01,ominimal}), we adopt an approach based on first-order formul\ae. 
However, in many cases, solutions, or approximated solutions, of the differential equations
are computed before the automaton analysis (see, e.g., \cite{ominimal}). 
Whenever such (approximated) solutions are 
polynomials, the same dynamics expressed by differential equations can be 
defined in Tarski's theory. 

Comparing our definition with the one in \cite{focore2008}, we can notice that we add the condition	
\emph{$Dyn(v)[p,q,0]$ implies $p=q$}. Intuitively, this means that if we are in $p$ at time $0$, 
we can reach a point different from $p$ through a continuous dynamics
only if we let time flow. 
This  assumption allows us to both get flow continuity at time $0$ 
and 
slightly simplify the reachability 
formul\ae\ with respect to the ones defined in \cite{focore2008}. 
\begin{example}\label{ex:HA}
			Figure~\ref{fig:HA} depicts a graphical representation of the hybrid automaton 
			$H = (Z,Z',T,\mathcal{V},\mathcal{E},Inv,$ $Dyn,Act,Res)$, where 
			$Z = \langle Z_1 \rangle$ and $Z' = \langle Z'_1 \rangle$ and both $Z_1$ and 
			$Z'_1$ are variables over $\mathbb{R}$;
			$\mathcal{V} = \{v\}$ and $\mathcal{E} = \{(v,v)\}$; 
			$Inv(v)[Z] \deq -100\leq Z_1 \leq 100$;
			$Dyn(v)[Z,Z',T] \deq (T=0\wedge Z_1'=Z_1)\vee(T>0 \wedge Z_1  <2* Z'_1 \leq 2*Z_1)$;
			$Res((v,v))[Z,Z'] \deq Z_1 < 2*Z'_1 < 2*Z_1$; 
			$Act((v,v))[Z] \deq \top$.
			
			\begin{figure}
			\begin{center}
				\begin{tikzpicture}[shorten >=1pt,node distance=2cm,auto] 

  \tikzstyle{every state}=[fill=black!20!white,draw,text=black]
				\node[state] (q_0) {$\substack{-100\leq Z_1 \leq 100 \\ \\ Dyn(v)[Z,Z',T] }$};
		
				\path[->] (q_0)	edge [loop right]	node	 {$\substack{ \frac{Z_1}{2} < Z'_1 < Z_1 }$} ();
		
				\end{tikzpicture} 
			\end{center}
			\caption{The hybrid automaton described in Example~\ref{ex:HA}.}
			\label{fig:HA}
			\end{figure}
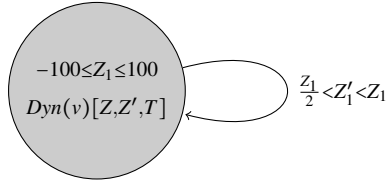

		\end{example}
			
\subsection{Standard Semantics}\label{sec:semantics}
	The formula $Dyn(v)[Z,Z',T]$ holds if there exists a continuous flow going 
	from $Z$ to $Z'$ in $T$ time-instants. We admit an infinite number of flows, which can 
	also be self-intersecting. Our semantics imposes the continuity of such flows.
	
	\begin{definition}[Hybrid Automata - Semantics]\label{sem_ha}
		A \emph{state} $\ell$ of $H$ is a pair $\langle v,r \rangle$, where $v \in \mathcal{V}$ is a location
		and $s = \langle s_1, \dots, s_{d(H)} \rangle \in \mathbb{R}^{d(H)}$ is an assignment of values
		for the variables of $Z$. A state $\langle v,s \rangle$ is \emph{admissible} if 
		$Inv(v)[s]$ is true. We have two kind of transitions:		
		\begin{itemize}
		\item the \emph{continuous transition relation} $\rightarrow_C$: \\
			$\langle v,s \rangle \rightarrow_C \langle v,r \rangle$
			$\iff$ 
			there exists $f : \mathbb{R}_{\geq 0} \rightarrow \mathbb{R}^{d(H)}$
			continuous function such that $s = f(0)$, there exists $t \geq 0$ such that $r = f(t)$, and for each
			$t' \in [0,t]$ 
			$Inv(v)[f(t')]$ and $Dyn(v)[s,f(t'),t']$ hold;
		
		\item the \emph{discrete transition relation} $\rightarrow_D$: \\
			$\langle v,s \rangle \rightarrow_D \langle u,r \rangle$
			$\iff$ 
			$( v,u ) \in \mathcal{E}$ 
			and both the formul\ae~$Act((v,u))[s]$ and $Res((v,u))[s,r]$ holds.
		\end{itemize}
	\end{definition}
	Combining continuous and discrete transitions, we introduce the notions of \emph{trace} and
	\emph{reachability}. A trace is a sequence of continuous and discrete transitions. A point $r$
	is reachable from a point $s$ if there is a trace starting from $s$ and ending in $r$. We use
	$\ell \rightarrow \ell'$ to denote that either $\ell \rightarrow_C \ell'$ or $\ell \rightarrow_D \ell'$.
	\begin{definition}[Hybrid Automata - Reachability]
		A \emph{trace} of length $n$ of $H$ is a sequence of admissible states $\ell_0,\ell_1,\dots,\ell_n$,
		with $n \in \mathbb N_{>0}$, such that:
		\begin{itemize}
			\item	for each $j\in [1,n]$ it holds $\ell_{j-1} \rightarrow \ell_j$;
			\item	for each $j\in [1,n-1]$ if $\ell_{j-1} \not\rightarrow_D \ell_{j}$, then $\ell_{j} \rightarrow_D \ell_{j+1}$.
		\end{itemize}		
		
		In $H$, $s \in \mathbb{R}^{d(H)}$ \emph{reaches} 
		$r \in \mathbb{R}^{d(H)}$ if there exists a trace $\ell_0,\dots,\ell_n$ of $H$ such that
		$\ell_0 = \langle v,s \rangle$ and $\ell_n = \langle u,r \rangle$, for some $v,u \in \mathcal{V}$.
		A set $\mathbb I\subseteq \mathbb{R}^{d(H)}$ \emph{reaches}  
		$\mathbb F\subseteq \mathbb{R}^{d(H)}$ if there exists $s\in \mathbb I$ which reaches $r\in \mathbb F$.
	\end{definition}
	We impose the condition that, in a trace, continuous transitions do not occur consecutively.
	In hybrid automata whose flows are solutions of autonomous differential equations, the
	continuous transition relation is transitive, hence if a trace contains a sequences of
	consecutive continuous transitions, it can be reduced to a trace without consecutive
	continuous transitions. However, Definition \ref{syn_ha} allows also automata whose 
	continuous transition relation is not transitive. 
	This can occur when the dynamics are solutions of non-autonomous differential equations. 
	For instance, if the formula based dynamics are $\langle X_0+T,Y_0+T^2\rangle$, 
	the set of points reachable from 
	$\langle 0,0 \rangle$ is, of course, $R=\{\langle t,t^2\rangle | t \in \mathbb{R}_{\geq 0} \}$. 
	However, since, for every 
	$r \in \mathbb{R}$, there exists a tuple $\langle t,r*t \rangle$ in $R$, by admitting multiple successive 
	continuous transitions we would have obtained $\mathbb{R}^2$ as the reachability set which 
	is visibly wrong. 

Let us consider the case of hybrid automata whose
dynamics are of the form $Dyn(v)[Z,Z',T]\stackrel{def}{=}Z' = f_v(Z,T)$, with
$f_v$ continuous (e.g., the case of solutions of vector fields). Let us call such automata \emph{functional automata}.
For functional automata it is easy to write a formula $R^i_H[Z,Z']$ which models the reachability with $i$ discrete transitions (see, e.g., \cite{focore2008}). 
%
Since, such formul\ae\ are first-order formul\ae\ whose satisfiability is decidable, we can summarize this result saying that 
for this class of automata reachability within a fixed number of discrete transitions is decidable. However, the possibility of characterizing reachability within a fixed number of discrete transitions through formul\ae\ does not imply the decidability of the reachability problem. As a matter of fact, even if the satisfiability of $R^{i}_H[Z,Z']$ is decidable, in order to solve the reachability problem we would need to test the satisfiability of an infinite number of formul\ae, i.e., one for each $i\in \mathbb N$.

If we do not impose any further condition on the dynamics, it is not possible to write a first-order formula representing a continuous transition.
As a matter of fact, our definition of $\rightarrow_C$ requires the existence of a continuous function $f$ which satisfies 
the constraints $Inv$ and $Dyn$, i.e., it requires to decide whether a selection problem has a solution (see Section \ref{sec_prel}). 
However, selection problems are neither expressible in our first-order language nor decidable in the general case. This is the main difference between our approach and the one presented in \cite{franzle99}, where the continuity of the continuous transitions (called $act$) is not imposed. 
		
In the next section we introduce a class of hybrid automata which generalizes functional automata still 
allowing to translate continuous transitions into formul\ae. 

\subsection{Michael's Form Automata}
In order to ensure the existence of a continuous function satisfying both invariant and dynamic constraints we need to check that such constraints meet the hypothesis of a selection theorem \cite{bookselection}. In particular, as in \cite{focore2008}, we consider a class
of automata based on Michael's selection result. The following definitions express Michael's hypothesis in the context of hybrid automata.

First we characterize the set valued map from which the continuous selection will take place.	
	\begin{definition}[$I^v_p$,$F^v_p$]
		Let $H$ be a hybrid automaton. Let $v$ be a location of $H$ and $p \in \mathbb{R}^{d(H)}$
		such that $Inv(v)[p]$ holds. $I^v_p$ is the interval of time instants satisfying the following:
		$\forall T \in I^v_p~\exists Z' (Dyn(v)[p,Z',T] \wedge Inv(v)[Z'])$. The time instant $0$ belongs 
		to $I^v_p$, and $I^v_p$ is maximal with respect to the first two requirements.
		
		The function $F^v_p : I^v_p \rightarrow \wp(\mathbb{R}^{d(H)})$ is defined as
			$F^v_p(T) = \{ q \mid Dyn(v)[p,q,T] \wedge Inv(v)[q]\}$.
	\end{definition}
Since Michael's theorem \cite{bookselection} requires lower semi-continuity, closedness, and $\alpha$-paraconvexity
(see Section \ref{sec_prel}), we obtain the following class of hybrid automata. 
	
	\begin{definition}[MF Automata \cite{focore2008}]\label{mf_ha}
		We say that a hybrid automaton $H$ is in \emph{Michael's Form}, or simply a \emph{MF} automaton,
		if for each state $v \in \mathcal{V}$ and for each point $p \in \mathbb{R}^{d(H)}$ such
		that $Inv(v)[p]$ holds, the function $F^v_p$ is lower semi-continuous, and for each
		$t \in I^v_p$ the set $F^v_p(t)$ is closed and $\alpha$-paraconvex.
	\end{definition}

As proved in~\cite{focore2008} if a hybrid automaton is in Michael's form continuous transitions 
can be characterized through a formula. As a consequence, reachability within a fixed number of discrete transitions can be mapped 
into a satisfiability problem over semi-algebraic formul\ae, as follows.

Let $H$ be a MF automaton, and $v \in \mathcal{V}$
	one of its locations. Consider the formula:
	\begin{equation*}
		Tp(v)[Z,T] \deq \forall T' (0 \leq T' \wedge T' \leq T \rightarrow \exists Z' (Dyn(v)[Z,Z',T'] \wedge Inv(v)[Z'])).
	\end{equation*}
	If $p$ satisfies $Inv(v)$, 
	 then it follows that
	$t \in I^v_p$ if and only if $Tp(v)[p,t]$ is true.
	
	\begin{definition}[MF Automata - Reachability Formula \cite{focore2008}]\label{mf_ha_reach}
			Let $H$ be a MF automaton. 
The formula $Reach^i_H(v,v')[Z,Z']$ is inductively defined as follows:
	\begin{equation*}
		Reach^0_H(v,v')[Z,Z'] 	\deq \begin{cases}
			\bot		& \text{if $v\neq v'$,} \\
			\begin{aligned}
			\exists T (T \geq 0 \wedge Dyn(v)[Z,Z',T] \wedge Tp(v)[Z,T])\wedge\\
			Inv(v)[Z] \wedge Inv(v)[Z']
			\end{aligned} & \text{otherwise.}
		\end{cases}
	\end{equation*}

	\begin{align*}			
	Reach^{i+1}_H(v,v')[Z,Z'] \deq  \bigvee_{\tilde{v} \in \mathcal{V}} ( \exists Z_1 \exists Z_2(&Reach^i_H(v,\tilde{v})[Z,Z_1] \wedge
							Act((\tilde{v},v'))[Z_1] \wedge \\
							& Res((\tilde{v},v'))[Z_1,Z_2]  \wedge
		 					 Reach^0_H(v',v')[Z_2,Z']  ) ).\\
	\end{align*}
		Moreover, we define the formul\ae:
		\begin{equation*}
		Reach^i_H[Z,Z'] \deq \bigvee_{v,v'\in\mathcal{V}}Reach^{i}_H(v,v')[Z,Z'] \hskip0.5cm\textrm{ and }\hskip0.5cm
		Reach^{\leq i}_H[Z,Z']  \deq \bigvee_{j=0}^{i}Reach^{j}_H[Z,Z'].
		\end{equation*}

	\end{definition}
		  

As immediate consequence of~\cite{focore2008}, 
the above defined formul\ae\ correctly characterize
the notion of reachability within a fixed number of discrete transitions.

\begin{lemma}\label{lemma1}
Let $H$ be a MF automaton. Let $\mathbb{I},\mathbb{F}\subseteq \mathbb{R}^{d(H)}$ be represented by the formul\ae~$I[Z],F[Z]$,  respectively.
The set $\mathbb{I}$ reaches the set $\mathbb{F}$ if and only if there exists $i\in\mathbb{N}$ such that the formula $Reach^{i}_H[Z,Z']\wedge I[Z]\wedge F[Z']$ is satisfiable.
\end{lemma}

In \cite{focore2008} it has also been shown that given a hybrid automaton $H$ it is possible to decide whether it is a MF automaton or not, again through satisfiability problems over semi-algebraic formul\ae.

		\begin{example}
			Let us consider the 
			Example~\ref{ex:HA}. It is immediate to see that $H$ is a MF automaton.

In this case $Tp(v)[Z,T]$ is the formula $-100\leq Z_1\leq 100$, and hence $Reach^0_H[Z,Z']$ becomes
$-100\leq Z_1\leq 100 \wedge$ $-100\leq Z'_1\leq 100 \wedge$ $\frac{Z_1}{2}<Z_1'\leq Z_1$.

			For this automaton, 
			the computation of the reachability set does
			not converge. More precisely, at each iteration, while the reach set upper bound remains constant,
			the lower bound decreases by a quarter with respect to the lower bound of the 
			previous reach set. Thus, only after an infinite number of iterations the lower bound would converge to zero.
		\end{example}


\section{Approximated Semantics}\label{sec_app}

The standard semantics $\ls\varphi\rs$ of a formula $\varphi$ with $n$ free-variables over the reals is a subset of $\mathbb{R}^n$.
Hence, once we have fixed a standard semantics function $\ls\cdot\rs$ which maps formul\ae~with $n$ free-variables into subsets of $\mathbb{R}^n$, every other function of this type can be seen as an approximated semantics.
\begin{definition}[Approximated Semantics]
An \emph{approximated semantics} is a function 
$|| \cdot ||:\bigcup_{n\in\mathbb N} \mathcal F_n \rightarrow \bigcup_{n\in\mathbb N} \wp(\mathbb R^n)$
such that for each $\varphi\in \mathcal F_n$, it holds that $|| \varphi||\subseteq \mathbb R^n$. 
\end{definition}
If $\lu \cdot \ru$  is an approximated semantics such that for each formula $\varphi$ it holds $\lu \varphi \ru\subseteq \ls \varphi \rs$, we can say that $\lu \cdot \ru$ is an \emph{under-approximation} of the standard semantics, or simply an \emph{under-approximation semantics}. Similarly, if $\lo \cdot \ro$ is such that $\ls \varphi \rs \subseteq \lo \varphi \ro$ is always true, then $\lo \cdot \ro$ is an \emph{over-approximation semantics}. There are approximated semantics which are neither under nor over-approximations.


In~\cite{DEDS09} the authors observe 
	that dense unbounded domains, which are the cause of undecidability of the reachability
	problem, are often abstractions of real world domains. In particular, they notice that,
	especially in the context of biological simulation, it is useful to avoid the ability to 
	distinguish between values whose distance is less than a fixed $\epsilon$.
	They introduce a new class of semantics for first-order formul\ae, called $\epsilon$-semantics, which guarantee
	the decidability of reachability in the case of automata with bounded invariants.
	\begin{definition}[$\epsilon$-Semantics \cite{DEDS09}]\label{def:eps_semantics}
		Let $\epsilon \in \mathbb R_{>0}$.
		For each formula $\psi\in\mathcal F_n$ let $\la \psi \ra_\epsilon \subseteq \mathbb{R}^n$,
		be such that:
		
		\begin{tabular}{l l}
			$(\epsilon)$	&	$\la \psi \ra_\epsilon = \emptyset$ or exists $p \in \mathbb{R}^n$ s.t. $B( p , \epsilon) \subseteq \la\psi\ra_\epsilon$  \\
			$(\cap)$		&	$\la \psi_1 \wedge \psi_2 \ra_\epsilon \subseteq \la \psi_1 \ra_\epsilon \cap \la \psi_2 \ra_\epsilon$ \\
			$(\cup)$		&	$\la \psi_1 \vee \psi_2 \ra_\epsilon = \la \psi_1 \ra_\epsilon \cup \la \psi_2 \ra_\epsilon$ \\
			$(\forall)$		&	$\la \forall X \psi[X,Z] |\}_\epsilon = \la \bigwedge_{r \in \mathbb{R}} \psi[r,Z] \ra_\epsilon$ \\
			$(\exists)$	&	$\la \exists X \psi[X,Z] \ra_\epsilon = \bigcup_{r \in \mathbb{R}} \la  \psi[r,Z] \ra_\epsilon$\\	
			$(\neg)$		&	$\la \psi \ra_\epsilon \cap \la \neg\psi \ra_\epsilon = \emptyset$	\\
		\end{tabular}
		
		Any semantics $\la \cdot \ra_\epsilon$ satisfying the above conditions is said
		to be an \emph{$\epsilon$-semantics}.
	\end{definition}
	In the above definition, as done in \cite{DEDS09},  with a slight abuse of notation we
	use $\bigwedge_{r\in \mathbb R}\psi$ to treat an infinite conjunction of formul\ae\ as a formula.
The Algorithm~\ref{galgo:reachability}, given in~\cite{DEDS09}, computes the sets of reachable states of a given automaton and 
describes sets of points through formul\ae. 

\begin{algorithm}\label{algo_reach}
		\caption{Reachability($H,I[Z],\la\cdot \ra_\epsilon$)}
		\label{galgo:reachability}
		\begin{algorithmic}[1]
		    \STATE $R[Z]\leftarrow I[Z]$
		    \STATE $N[Z]\leftarrow \bot$
			\REPEAT 
			\STATE $R[Z]\leftarrow R[Z]\vee N[Z]$
			\STATE $N[Z]\leftarrow \exists Z' (Reach^{\leq 1}_H[Z',Z]\wedge R[Z'])$
			\UNTIL {$\la N[Z]\wedge \neg R[Z] \ra_\epsilon \neq \emptyset$ is true} 
			\RETURN $\la R[Z] \ra_\epsilon$
		\end{algorithmic}
	\end{algorithm}

The reachability is computed incrementing at each step the number of 
discrete transitions: new reachable
sets of points are computed until they become too small to be identified in the $\epsilon$-semantics.
In the case of hybrid automata with bounded invariants such reachability algorithm always terminates. 
If we replace the $\epsilon$-semantics with the standard one, each step of the above algorithm under-approximates the set of reachable points.
However, it still may not end even in the case of bounded invariants.

The \emph{sphere semantics} $\lsph \cdot \rsph_\epsilon$ (see e.g.,~\cite{DEDS09}) is an $\epsilon$-se\-man\-tics which 
 is neither an over nor an under-approximation semantics. 
%
		The set $\lsph \psi \rsph_\epsilon$, where $\epsilon \in\mathbb R_{>0}$, is defined 
		as follows:
		
		\begin{tabular}{l l}
			$(\epsilon)$	&	$\lsph t_1 \circ t_2 \rsph_\epsilon \deq B(\ls t_1 \circ t_2 \rs, \epsilon)$, for $\circ \in \{ =, <\}$  \\
			$(\cap)$		&	$\lsph \psi_1 \wedge \psi_2 \rsph_\epsilon \deq \bigcup_{B( p, \epsilon)  \subseteq \lsph \psi_1 \rsph_\epsilon \cap \lsph \psi_2 \rsph_\epsilon } B(p, \epsilon)$  \\
			$(\cup)$		&	$\lsph \psi_1 \vee \psi_2 \rsph_\epsilon \deq \lsph \psi_1 \rsph_\epsilon \cup \lsph \psi_2 \rsph_\epsilon$ \\
			$(\forall)$		&	$\lsph \forall X \psi[X,Z] \rsph_\epsilon \deq \bigcup_{B(p, \epsilon)  \subseteq \bigcap_{r\in \mathbb R}\lsph \psi[r,Z]\rsph_\epsilon} B({p},\epsilon)$ \\
			$(\exists)$	&	$\lsph \exists X \psi[X,Z] \rsph_\epsilon \deq \bigcup_{r \in \mathbb{R}} \lsph  \psi [r,Z] \rsph_\epsilon$\\	
			$(\neg)$		&	$\lsph \neg \psi \rsph_\epsilon \deq \bigcup_{B( p, \epsilon) \cap \lsph \psi \rsph_\epsilon = \emptyset} B(p, \epsilon)$	\\
		\end{tabular}
		
	\begin{example}
		Let $H$ be as in Example~\ref{ex:HA}.
		The sphere semantics with $\epsilon = 0.5$  gives us 
		$\lsph Reach^0_H[10,Z'] \rsph_\epsilon = (4.5,10.5)$,
		$\lsph Reach^1_H[10,Z'] \rsph_\epsilon = (0.75,10.5)$,
		$\lsph Reach^2_H$ $[10,Z'] \rsph_\epsilon$ $= (-0.19,10.5)$, and
		$\lsph Reach^3_H[10,$ $Z'] \rsph_\epsilon = (-0.42,10.5)$.
		Thus, the reachability algorithm described in~\cite{DEDS09} over $H$, instantiated with the
		sphere semantics with $\epsilon = 0.5$, halts and returns as result the set $(-0.19,10.5)$, since the
		difference between $\lsph Reach^2_H[10,Z'] \rsph_\epsilon$ and 
		$\lsph Reach^3_H[10,Z'] \rsph_\epsilon$
		is smaller than an open sphere of radius $\epsilon = 0.5$.
		
		This example also points out that
	whenever a variable is quantified in a formula, the $\epsilon$-semantics evaluates it with all the possible constants, and hence there are no approximation effects on it. As a matter of fact $\lsph Reach^1_H[10,Z']\rsph_\epsilon$ does not include the interval $[10.5,11,5)$ which would be included if the quantified variables in the formula $Reach^1_H[10,Z']$ were over-approximated.
	\end{example}

It is immediate to prove that, due to rule $(\neg)$, $\epsilon$-semantics are never over-approximation semantics.
As a consequence of this fact and of the structure of the reachability algorithm, the approach proposed in \cite{DEDS09} is not taylored for over-approximating reachability. 
In Section~\ref{sec:bottom} we show how to exploit $\epsilon$-semantics for under-approximation.
		
\section{Disturbed Automata}\label{sec_dist}

	In this section we present the approach proposed by Fr\"anzle in~\cite{franzle99}
	for the over-approximation of reachability. In particular, we 
	briefly recall the framework described in \cite{franzle99} and then we establish some general relationships 
	with the framework we introduced in Section \ref{sec_ha}. 
	
	Fr\"anzle
	noticed that real hybrid systems are always subject to noise, suspecting that their continuous
	components can provide only finite memory. If so, the state space of such automata would
	be the product of the size of the discrete state space and the effective size of the continuous
	state space modulo noise. This means that the reach set computation of hybrid automata modeling
	real systems, should converge finitely, yielding decidability of state reachability.
	 
	The definition of hybrid automata given in \cite{franzle99} slightly differs from 
	Definition~\ref{syn_ha}. Specifically, activations and resets are characterized
	by a single formula called \emph{transition predicate} $trans_{v\rightarrow v'}$. Similarly, invariants and dynamic laws
	are merged in the \emph{activity predicate} $act_{v}$, which does not impose any constraint on the
	continuity of the dynamic laws.
	
		\begin{definition}[Fr\"anzle Hybrid Automata - Syntax~\cite{franzle99}]\label{def:franzle_HA}
		A \emph{Fr\"anzle hybrid automaton} 
			$H = (\mathcal V,Z',
			(act_{v})_{v \in \mathcal V},$
			$(trans_{v\rightarrow v'})_{v,v' \in \mathcal V},
			(initial_v)_{v \in \mathcal V},
			(safe_v)_{v \in \mathcal V})$
		of dimension $d \in \mathbb{N}$ consists of the following components:
		\begin{itemize}
			\item	$\mathcal V$ is a finite set, representing the discrete locations;
			\item $Z$ is a vector of variable names of dimension $d$, 
				representing the continuous variables of $H$;
			\item each $v\in \mathcal V$ is labeled with a formula $act_v[Z,Z']$ representing the
				continuous activities and corresponding state constraints;
			\item each pair of locations $v,v'\in \mathcal V$ is labeled with a formula $trans_{v\rightarrow v'}[Z,Z']$
				representing the
				discrete transitions and their guarding conditions;
			\item each $v\in \mathcal V$ is labeled with the formul\ae\	$initial_v[Z]$ and $safe_v[Z]$
				representing the initial and the safe states of the hybrid automaton.
		\end{itemize}
	\end{definition}
	
	As far as the semantics is concerned, the reachability formula  $\Phi(H)^{i}_{v\rightarrow v'}$
	is defined as follows~\cite{franzle99}:
	%
	\begin{equation*}
		\Phi(H)^0_{v\rightarrow v'}[Z,Z']\deq
		\begin{cases}
			act_v[Z,Z']		&	\text{if $v = v'$,} \\
			\bot				&	\text{otherwise.} \\
		\end{cases}
	\end{equation*}

	\begin{equation*}
    			\Phi(H)^{i+1}_{v\rightarrow v'}[Z,Z'] \deq\   \bigvee_{\tilde{v}\in\mathcal V} \exists Z_1\exists Z_2(\Phi(H)^{i}_{v\rightarrow \tilde{v}}[Z,Z_1]\wedge 
			 trans_{\tilde{v}\rightarrow v'}[Z_1,Z_2]\wedge act_{v'}[Z_2,Z']). \\
	\end{equation*}

An automaton is said to be \emph{safe} if the initial states can only reach \emph{safe} states. 

We formalize how a MF automaton can be mapped into an automaton 
w.r.t.~Definition~\ref{def:franzle_HA}.

\begin{definition}[Corresponding Fr\"anzle Automaton]
	Given a MF automaton $H=(Z,Z',T,\mathcal V,\mathcal E,$ $Inv,Dyn,$ $Act,Res)$ and two formul\ae\ $I[Z]$ and $F[Z]$,  
	the \emph{corresponding Fr\"anzle automaton} is the automaton 
$Fr(H,I,F) \deq (\mathcal V,$ $Z', (act_v$ $)_{v\in\mathcal V},$ $(trans_{v\rightarrow v'})_{v,v'\in\mathcal V},$ $(initial_v)_{v\in \mathcal V},$ $(safe_v)_{v\in\mathcal V})$ 
where:\begin{itemize}
	\item for each $v\in \mathcal V$, $act_{v}\deq Reach_H^0(v,v)$, $initial_v\deq Inv(v)[Z]\wedge I[Z]$, and $safe_v\deq Inv(v)[Z]\wedge \neg F[Z]$;
    \item for each $e=(v,v')\in \mathcal E$, $trans_{v\rightarrow v'}\deq Act(e)\wedge Res(e)$, while
    for each $(v,v')\not\in \mathcal E$, $trans_{v\rightarrow v'}\deq \bot$.
	\end{itemize}
\end{definition}

Such translation establishes the following relationship between the formul\ae\ $\Phi(Fr(H,I,F))^i_{v\rightarrow v'}$ of \cite{franzle99} and our reachability formul\ae.
	\begin{lemma}\label{lemma:phi_reach}
	Let $H$ be a MF automaton, $I[Z]$ and $F[Z]$ be two formul\ae. Let $Fr(H,I,F)$ be the corresponding Fr\"anzle automaton. For each $i\in \mathbb N$ it holds that
	$\ls \Phi(Fr(H,I,F))^i_{v\rightarrow v'}[Z,Z']\rs = \ls Reach_H^i(v,v')[Z,Z']\rs$.
	\end{lemma}	

Hence, our notion of reachability corresponds to a non safety condition.

\begin{theorem}\label{theo:fr}
	Let $H$ be a MF automaton, $I[Z]$ and $F[Z]$ be two formul\ae\ representing $\mathbb I$ and $\mathbb F$, respectively. Let 
	$Fr(H,I,F)$ be the corresponding Fr\"anzle automaton.
	$\mathbb I$ re\-a\-ches $\mathbb F$ in $H$ if and only if $Fr(H,I,F)$ is not safe.
\end{theorem}	
	Given a hybrid automaton $H$, Fr\"anzle defines $\widetilde{H}$,
	called the \emph{disturbed variant of noise level $\epsilon$}, as the automaton obtained 
	from $H$ perturbing all the activity predicates, i.e., expanding the activity predicates 
	by an open sphere of radius $\epsilon$. Thus, since every activity predicate of an automaton
	$Fr(H,I,F)$ corresponds to the reachability formula $Reach^0_H$,
	we can define a disturbed variant of $H$ as follows.

    \begin{definition}[Disturbed Automata]
	Let us consider a MF automaton $H$. 
	The MF automaton $\widetilde{H}=(Z,Z',T, \mathcal{V},\mathcal E,$ $\widetilde{Inv},\widetilde{Dyn},Act,Res)$ is a \emph{disturbed variant} of $H$ if and only if for each $v\in \mathcal V$ it holds that
	$\ls Reach_H^0(v,v)[Z,Z']\rs\subseteq \ls Reach_{\widetilde{H}}^0$ $(v,v)[Z,Z']\rs$.
	
	Moreover, let $\epsilon\in\mathbb R_{>0}$. We say that a disturbance $\widetilde{H}$ of $H$ is a \emph{disturbance of noise level $\epsilon$ or more} if and only if for each $v\in\mathcal V$ it holds that 
	$
		\ls \exists Z''(Reach_H^0(v,v)[Z,Z'']\wedge \delta(Z'',Z')<\epsilon)\rs\subseteq 
		\ls Reach_{\widetilde{H}}^0(v,v)[Z,Z']\rs
	$.	
\end{definition}	
In the above definition we refer to the standard euclidian distance $\delta$. 
Our definition of disturbed variants is an instance of Fr\"anzle definition in the following sense.
\begin{lemma}\label{lemma:correspond}
Let $H$ be a MF automaton. $I[Z]$ and $F[Z]$ be two formul\ae. Let $Fr(H,I,F)$ be the corresponding Fr\"anzle automaton.
If $\widetilde{H}$ is a disturbed variant of $H$, $Fr(\widetilde{H},I,F)$ is a disturbed variant of $Fr(H,I,F)$ 
w.r.t.~Fr\"anzle's definition.
If $\widetilde{H}$ is a disturbance of noise level $\epsilon$ or more, 
so is $Fr(\widetilde{H},I,F)$.
\end{lemma}

As a consequence, exploiting Lemma 2 of \cite{franzle99}, we get the following theorem which states how $\widetilde{H}$ can
be used to over-approximate reachability over $H$.
\begin{theorem}\label{theo:union}
Let $H$ be a MF automaton, $\widetilde{H}$ be a disturbed variant of $H$ of noise level $\gamma>0$, and $I[Z]$ be a formula. If 
$\ls \widetilde{Inv}(v)[Z]\rs$ is bounded for each $v\in \mathcal V$, then there exists $i\in \mathbb N$ such that:  
$$\bigcup_{n\in \mathbb{N}}\ls 
Reach_H^n
[Z,Z']\wedge I[Z]\rs\subseteq \bigcup_{n=0}^i
\ls
Reach_{\widetilde{H}}^n
[Z,Z'] \wedge I[Z]\rs.$$
Moreover, $i$ can be effectively computed.
\end{theorem}

\section{Mixing the Approaches}\label{mixing}
In this section we first re-describe Fr\"anzle's reachability algorithm in terms of approximated semantics, obtaining an over-approximation reachability algorithm which does not explicitly refer to $\tilde{H}$. Then we focus on under-approximations of reachability based on $\epsilon$-semantics. 
\subsection{Over-Approximation}\label{sec:over}
	We define a new approximated semantics, named \emph{tilde semantics}, which captures the introduction of noise in hybrid automata. 
	

\begin{definition}[Tilde Semantics]
Let $\psi$ be a formula and let $\epsilon \in \mathbb R_{>0}$. The \emph{tilde semantics} of $\psi$ is 
$\lt \psi \rt_\epsilon \deq B(\ls  \psi \rs , \epsilon)$.
\end{definition}

Such semantics applied to $H$ under-approximates each $\epsilon$-disturbance of $H$ in the following sense.

\begin{theorem}\label{teo:tilde_in_noise}
Let $H$ be a MF automaton and $\widetilde{H}$ be a disturbance of noise level $\epsilon$ or more.
For each $v,v'\in \mathcal V$, for each $p\in \mathbb R^{d(H)}$ it holds that
$\lt Reach_H^i(v,v')[p,Z'] \rt_\epsilon \subseteq \ls Reach_{\widetilde{H}}^i(v,v')[p,Z']\rs$.
\end{theorem}
Hence, exploiting Theorem \ref{teo:tilde_in_noise} we get the following result.
\begin{corollary}\label{cor:tilde_noise}
Let $H$ be a MF automaton, $\widetilde{H}$ be a disturbance of noise level $\epsilon$ or more with respect to $\delta$, and $I[Z]$ be a formula.
If for each $v\in \mathcal V$ it holds that 
$\ls \widetilde{Inv}(v)[Z]\rs$ is bounded, then there exists $i\in \mathbb N$ such that 
	$\bigcup_{n\in \mathbb{N}}\ls Reach_H^n [Z,Z']\wedge I[Z]\rs 
	\subseteq
	\bigcup_{n\in \mathbb{N}}\lt Reach_H^n [Z,Z'] \wedge I[Z]\rt_\epsilon
	\subseteq
	\bigcup_{n=0}^i \ls Reach_{\widetilde{H}}^n [Z,Z'] \wedge I[Z]\rs$. 
%
Moreover, $i$ can be effectively computed.
\end{corollary}

The following definition characterizes an $\epsilon$-disturbance 
whose semantics is minimal, i.e., it is included in all $\epsilon$-disturbance semantics.

\begin{definition}[Tilde Transformation]\label{def:computing_tilde}
	Let $\mathcal{T}$ be a first-order theory over the reals, $\psi[Z]$ be any first-order formula 
	$\mathcal{T}$-definable, and $\epsilon \in \mathbb{R}_{>0}$. The \emph{tilde transformation} 
	of $\psi[Z]$ is defined as follows:
	$$\widetilde{(\psi[Z])}_\epsilon \deq \exists Z_0(\psi[Z_0] \wedge \delta(Z_0,Z) < \epsilon).$$
\end{definition}

\begin{theorem}\label{theo:boh}
	Let $\mathcal{T}$ be any first-order theory and 
	$\psi[X] \in \mathcal{T}$.  
	The tilde semantics of $\psi[X]$ is $\mathcal{T}$-definable 
	and, in particular, $\lt\psi[X]\rt_\epsilon = \ls \widetilde{(\psi[X])}_\epsilon\rs$
	for all $\epsilon \in \mathbb{R}_{>0}$.
\end{theorem}
%

\begin{definition}[Minimum Disturbed Variant]
Let $H$ be a MF automaton, $I[Z]$ and $F[Z]$ be formul\ae, and $\epsilon\in \mathbb R_{>0}$. The \emph{minimum $\epsilon$ disturbed variant of $H$}, 
	$\widetilde{Fr(H,I,F)}=(\mathcal V,Z',$ $(\widetilde{act}_{v})_{v \in \mathcal V},
	(\widetilde{trans}_{v\rightarrow v'})_{v,v' \in \mathcal V},$
	$(\widetilde{initial}_v)_{v \in \mathcal V},
	(\widetilde{safe}_v)_{v \in \mathcal V})$, is the disturbed variant of $Fr(H,I,F)$ of noise level $\epsilon$ obtained considering
for each $v\in \mathcal V$, $\widetilde{act}_{v}\deq \exists Z''(Reach_H^0(v,v)[Z,Z'']$ $\land \delta(Z'',Z')<\epsilon)$, 
while the other components are defined as for $Fr(H,I,F)$.

\end{definition}

Tilde semantics precisely captures the continuous semantics of $\widetilde{Fr(H,I,F)}$.

\begin{lemma}\label{lem:tilde_overline}
Let $H$ be a MF automaton, $I[Z]$ and $F[Z]$ be formul\ae, and $\epsilon\in \mathbb R_{>0}$. $\widetilde{Fr(H,I,F)}$ is an $\epsilon$ disturbed variant of $Fr(H,I,F)$. Moreover, for each $v\in \mathcal V$ and $p\in \mathbb R^{d(H)}$ it holds
$\lt Reach_H^0(v,v)[p,Z'] \rt_\epsilon = $ $\ls \Phi(\widetilde{Fr(H,I,F)})^0_{v \rightarrow v}[p,Z']\rs$.
\end{lemma}
%
The above result cannot be generalized to $Reach_H^n$ and
$\Phi(\widetilde{Fr(H,I,F}))^n$. In particular, the tilde semantics of the first-one in the general case is strictly included in the standard semantics of the second one. This is due to the fact that 
the first formula is built closing intermediate steps through quantifiers, which means that the intermediate steps are not approximated. On the other hand, in the second formula each step is over-approximated.

Lemma \ref{lem:tilde_overline} enables us to rephrase the algorithm described by Fr\"anzle as Algorithm~\ref{talgo:reachability}. 

	\begin{algorithm}\label{fr_reach}
		\caption{Tilde($H,I[Z],\epsilon$)}
		\label{talgo:reachability}
		\begin{algorithmic}[1]
		    \STATE $R\leftarrow \bigcup_{p\in \ls I[Z]\rs} \lt Reach_H^{0}[p,Z']\rt_{\epsilon}$
			\REPEAT 
			\STATE $V\leftarrow R$
			\STATE $\begin{aligned}R\leftarrow\bigcup_{p\in R}\ls \bigvee_{(v,v')\in \mathcal E}(Act((v,v'))[p]\land  Res((v,v'))[p,Z']\rs\end{aligned}$
			\STATE $R\leftarrow \bigcup_{p\in R} \lt Reach_H^{0}[p,Z']\rt_\epsilon$
			\STATE $R\leftarrow R\cup V$
			\UNTIL {$\bigcup_{p\in V}\ls Reach^{\leq 1}_H[p,Z']\rs\not\subseteq V$ is true} 
			\RETURN $V$
		\end{algorithmic}
	\end{algorithm}

\begin{theorem}\label{corr_fr}
Let $H$ be a MF automaton with bounded invariants, $I[Z]$ be a formula, and $\epsilon\in \mathbb R_{>0}$.
\emph{Tilde($H,I[Z],\epsilon$)} always terminates returning a set $R$ such that 
$\bigcup_{n\in \mathbb{N}}\ls Reach_H^n[Z,Z']\wedge I[Z]\rs\subseteq R$, i.e., it over-approximate reachability.
\end{theorem}	


		\begin{example}
			Let us consider the hybrid automaton $H$ described by Example~\ref{ex:HA}.
			The sets $R$ and $V$ calculated by the first three iterations of Algorithm~\ref{talgo:reachability},
			with $\ls I[Z]\rs  = \{10\}$ and $\epsilon = 0.5$, are 
			$R^1 = (4.5,10.5)$ and $V^1 = (1.13,10.5)$, $R^2 = (0.75,10.5)$ and $V^2 = (0.19,10.5)$,
			$R^3 = (-0.19,10.5)$ and $V^3 = (-0.05,10.5)$. Since $V^3 \subset R^3$, the algorithm
			halts and returns as result the set $(-0.19,10.5)$, which is an over-ap\-prox\-ima\-tion
			of the standard reach set of $H$, i.e., $(0,10]$.
		\end{example}

\subsection{Under-Approximation}\label{sec:bottom}

Fr\"anzle's approach can be used to under-approximate reachability over $H$ by defining an automaton $H'$ such that $\widetilde{H'}=H$. 
However, this would give us an under-approximation algorithm in which at each step an under-approximation of $Reach^1$ is applied.
In this section 
we show that the approach proposed in \cite{DEDS09} can always be used to under-approximate reachability, no matter which $\epsilon$-semantics is considered. Moreover, when the considered $\epsilon$-semantics is an under-approximation semantics, we get an algorithm in which the same under-approximations are applied to both termination conditions and output.

We start introducing a new semantics, called \emph{bottom semantics}.
\begin{definition}\label{def:bottom_semantics}
		Let $\psi$ be a formula and $\epsilon\in\mathbb R_{> 0}$.
		The set $\lb \psi \rb_\epsilon$ is the \emph{bottom semantics} of $\psi$ and it is defined by structural induction on $\psi$ itself as follows:
		\begin{itemize}
			\item	$\lb t_1 \circ t_2 \rb_\epsilon = \bigcup_{B(p,\epsilon) \subseteq \ls t_1 \circ t_2\rs } B(p,\epsilon)$,
				for $\circ \in \{ =, < \}$;
			\item	$\lb \psi_1 \wedge \psi_2 \rb_\epsilon = \bigcup_{B(p,\epsilon) \subseteq \lb \psi_1 \rb_\epsilon \cap \lb \psi_2 \rb_\epsilon} B(p,\epsilon)$;
			\item	$\lb \psi_1 \vee \psi_2 \rb_\epsilon = \lb \psi_1 \rb_\epsilon \cup \lb \psi_2 \rb_\epsilon$;
			\item	$\lb \forall X \psi[X,Z] \rb_\epsilon = \bigcup_{B(p,\epsilon) \subseteq \bigcap_{r \in \mathbb{R}} \lb \psi[r,Z] \rb_\epsilon } B(p,\epsilon)$;
			\item	$\lb \exists X \psi[X,Z] \rb_\epsilon = \bigcup_{r \in \mathbb{R}} \lb \psi[r,Z] \rb_\epsilon$;
			\item	$\lb \neg\psi \rb_\epsilon = \bigcup_{B(p,\epsilon) \cap \ls \psi\rs  = \emptyset} B(p,\epsilon)$.
		\end{itemize}
	\end{definition}

The bottom semantics is an $\epsilon$-semantics. Moreover, any variable assignment, that satisfies a formula $\psi$ in the bottom semantics, satisfies $\psi$ in the standard Tarski's semantics too.
	
\begin{lemma}\label{lemma_under}
The bottom semantics is an $\epsilon$-semantics.
Moreover, $\lb \psi \rb_\epsilon \subseteq \ls \psi \rs$ for each formula $\psi$.
\end{lemma}

The bottom semantics is 
definable in the Tarski's theory, i.e., 
if $\psi$ is a formula of the first-order language 
of the reals equipped of sum, product and comparison relations, then 
there exists a formula $\widehat{\psi}_\epsilon$ such that 
$\lb\psi\rb_\epsilon = \ls\widehat{\psi}_\epsilon\rs$. Moreover, 
$\widehat{\psi}_\epsilon$ is computable.

\section{Implementation and Tests}\label{case}

We implemented our framework and algorithms exploiting the translations of approximated semantics into the standard one, and then relying on quantifier elimination packages.
In particular, we developed a 
tool to work with Tarski's formul\ae\ and to compute,  
given a formula $\psi$, the formul\ae\ $\widetilde{\psi}_\epsilon$ and $\widehat{\psi}_\epsilon$.
The evaluation of the termination conditions of both the algorithms are obtained by first 
translating the $\epsilon$-semantics of the involved formul\ae\ in the corresponding 
Tarski's formul\ae, and then using \texttt{Redlog}-\texttt{QEPCAD} 
to eliminate the quantifiers 
as described in~\cite{Sturm:2011:VSU:1993886.1993935}. 

\begin{figure}
		\begin{center}
		\begin{tikzpicture}[scale=1.5,domain=0:4,fill=blue!20]  
	
			\draw[dashed,fill=lightgray!20] (-0.65,-0.5) rectangle (0.65,0.5) node[below=2]{}; 
			\draw (0,-0.5) node [above right] {$-y_{u}$};		
			\draw (-0.65,0) node [above right] {$-x_{u}$};
			
			\draw[->] (-3,0) -- (1.5,0) node[right] {$x$}; 					
			\draw[->] (0,-2.2) -- (0,1.5) node[above] {$y$}; 
			\draw[very thin,color=lightgray] (-2.8,-1.9) grid (1.49,1.49); 		
		
			\draw[fill=black] (0,-1.8) circle (0.075) node[right] { $Car$};		
			\draw[fill=black] (-2,0) circle (0.075) node[below=2] { $Train$};		
		
			\draw[domain=0.4:-0.4, color=black, dashed] plot(\x,{-0.8}) node[left] {$S$};	
			\draw[<->] (-2,0.7) -- node[above] {$Z_t$} (0,0.7);	
			\draw[<->] (0.9,-1.8) -- node[right] {$Z_c$} (0.9,0); 	
		
			\draw[->, very thick] (-2,0) --  node[above] {$V_t$} (-1.15,0); 	
			\draw[->, very thick] (0,-1.8) -- node[right] {$V_c$} (0,-1.05);  	

		\end{tikzpicture}
		\end{center}
		\caption{A case of study: a railroad crossing.}
		\label{fig:rail_crossing}
	\end{figure}
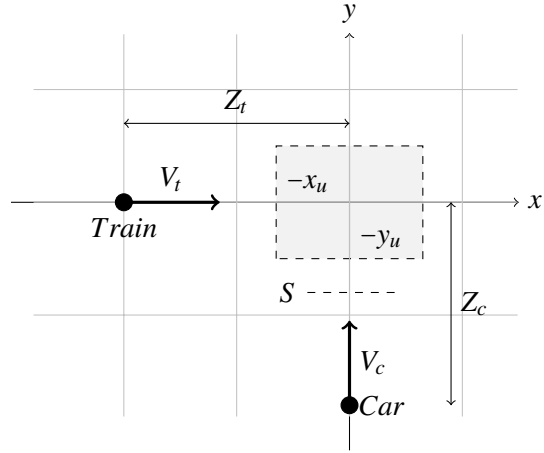

We tested our implementation on 
 a railroad crossing scenario without barriers (see Figure~\ref{fig:rail_crossing}): 
	a train and a car are simultaneously approaching to 
	the railroad crossing at coordinates $(0,0)$; 
	the train is moving along the $x$ axis with speed $V_t$ while the car has speed $V_c$ and 
	runs the $y$ axis. The variables $Z_t$ and $Z_c$ denote the
	distances between the railroad crossing and the train and between the railroad crossing and the car, respectively. 
	The car sensors can identify 
	the approaching train only above the line $S$. At that point it can decide to either 
	accelerate or slow down. The car acceleration $A_c$ should be once and for all and cannot be changed 
	anymore. 
	Our goal is to select an $A_c$ such that the two vehicles pass safely through the railroad crossing.
	
	In this case-study, we suppose we do not know with absolute precision the speeds of the two 
	vehicles. This means that, in a specific instant, the velocities belong to an interval,
	rather than being a single value. For this reason, we will use inequalities to describe the
	dynamic laws. 
	
	We model the scenario depicted in Figure~\ref{fig:rail_crossing} by a hybrid automaton 
	$H = (Z,Z',T,\mathcal{V},\mathcal{E},Inv,Dyn,$ $Act,Res)$ where: 
	\begin{itemize}
		\item $Z = \langle Z_t,Z_c,V_t, V_c, A_c \rangle$ and $Z' = \langle Z'_t,Z'_c,V'_t, V'_c, A'_c \rangle$ are variables over $\mathbb{R}^5$;
		\item $\mathcal{V} = \{q_0, q_c, q_s, q_u\}$ and $\mathcal{E} = \{e_0=(q_0,q_c),e_1=(q_c,q_u),e_2=(q_c,q_s) \}$; 
		\item 
		         $Inv(q_0) \deq Z_c\leq S \land  V_c \in [c_{m},c_{M}]  \land  V_t \in [t_{m},t_{M}]$; \\
			$Inv(q_c) \deq ((Z_c \in[S, y_{u}] \land Z_t\leq -x_{u}) \lor (Z_c \in [S, -y_{u}] \land Z_t\in [-x_{u}, x_{u}])) \land  V_c \in [c_{m},c_{M}]  
			\land  V_t \in [t_{m},t_{M}] $;\\
			$Inv(q_u) \deq Z_c \in [-y_{u}, y_{u}] \land Z_t\in [-x_{u},x_{u}]\land  V_c \in [c_{m},$ $c_{M}]  
			\land  V_t \in [t_{m},t_{M}]$;\\
			$Inv(q_s) \deq (Z_c \geq y_{u} \lor Z_t \geq x_{u}) \land  V_c \in [c_{m},c_{M}]  \land  V_t \in [t_{m},t_{M}] $;
		\item 
			$Dyn(q_c) \deq V_c' - A_c * T - V_c \in [-d,d] \land 2 * Z_c' - A_c * T^2 - 2 * V_c * T - 2 * Z_c \in [-d,d] \land Z_t' - V_t * T - Z_t \in [-d,d]$; \\ $Dyn(q_u)\deq Dyn(q_s) \deq$ $Dyn(q_0)[Z,Z',T]  \deq Z_t'-Z_t \in [t_{m},t_{M}] * T \land Z_c'-Z_c \in [c_{m}, c_{M}] * T$;
		\item $Act(e_0) \deq Z_c = S$; 
			$Act(e_1) \deq (Z_c \geq -y_{u}-d \land Z_t \in [-x_{u},x_{u}]) \lor (Z_t \geq -x_{u}-d \land Z_c \in [-y_{u},y_{u}])$; \\
			$Act(e_2) \deq  (Z_c = y_{u} \land Z_t < -x_{u}) \lor (Z_t = x_{u} \land Z_c <-y_{u})$.
		\item  $Res(e_2) \deq Res(e_1) \deq  A_c' \in [E_{m},E_M] \land V_c'\in[c_{m},c_{M}]   \land V_t'\in [t_{m},t_{M}]$;  
			$Res(e_0) \deq A_c' \in [E_{m},E_{M}]$.
	\end{itemize}
	where $c_m=1$ and $C_m=3$ are the minimal and the maximal admitted speed for the car,  
	$t_m=1$ and $T_m=3$ are the minimal and the maximal admitted speed for the train, and 
	$E_{m}=0$ and $E_{M}=10$ are the minimal and the maximal admitted car acceleration, respectively.

	Location $q_0$ corresponds to the phase in which the train and the car are far apart from the crossing.
	When the car reaches the $S$ line, the automaton goes to location $q_c$ and the car non-deterministically chooses 
	its acceleration. Whenever the car and the train cross the intersection at the same time, a collision occurs and the automaton 
	goes to the location $q_u$. The $q_s$ represents the safe situation in which at least one of the two vehicles has passed through the crossing, while 
	the other has not yet reached it.
	
%
%

We would like to decide, once the car has reached the position $S$, which accelerations avoid
the collision as a function of train speed, train position, and car speed.


\begin{figure}[!ht]
\begin{center}
\includegraphics[width=0.45\textwidth]{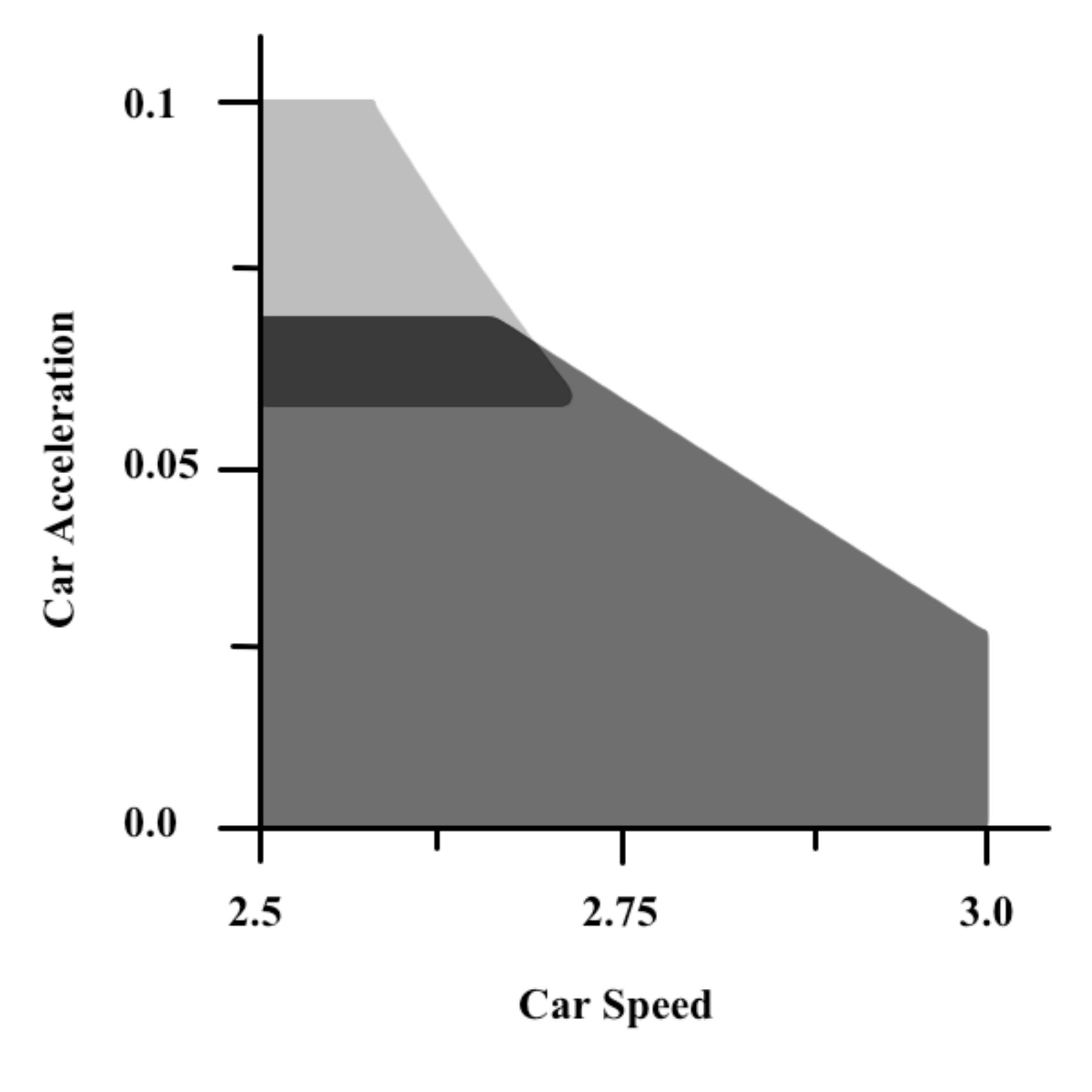}
\caption{A graphical representation of the analysis performed on  
the test case with $y_u=2$, $x_u=4$, and $d=0.2$. 
This figure depicts the space plane 
defined by the values $V_t=2$ and $Z_t=-20$ in the 
region $V_c\in [2.5,3.0]$ and $A_c \in [0,0.1]$.
The light, medium, and dark gray represent 
the sets $\lb \psi_{\textrm{sf}} \rb_\epsilon \setminus
\lt \psi_{\textrm{un}} \rt_\epsilon$, $\lb \psi_{\textrm{sf}} \rb_\epsilon \cap
\lt \psi_{\textrm{un}} \rt_\epsilon$, and 
$\lt \psi_{\textrm{un}} \rt_\epsilon \setminus \lb \psi_{\textrm{sf}} \rb_\epsilon$, respectively. 
If, given a car speed, we select an opportune 
acceleration such that the corresponding point is light gray colored, then 
we will certainly avoid the collision.}\label{fig:analysis}
\end{center}
\end{figure}

We modeled the automaton that represents the railroad 
crossing and 
we computed the two formul\ae\ $\psi_{\textrm{un}}$ and $\psi_{\textrm{sf}}$: 
the former represents all the situations in which the car 
has both reached the line $S$ and selected an acceleration $A_c$ that, 
sooner or later, leads to a collision; the latter characterizes all the states that 
avoid the collision itself.  
The formula $\psi_{\textrm{un}}$ depicts a jump 
over the edge $(q_0,q_c)$ and a successive continuous evolution 
ending up into the activation region of $(q_c,q_u)$, while 
$\psi_{\textrm{sf}}$ concludes the evolution into the activation of  $(q_c,q_s)$. 
We symbolically evaluated the tilde semantics of $\psi_{\textrm{un}}$ and the 
bottom semantics of $\psi_{\textrm{sf}}$ obtaining unquantified formul\ae, $\psi_{\textrm{un}}'$  
and $\psi_{\textrm{sf}}'$, respectively, in 4 free variables 
which represent the train speed, the train position, the car speed, and the car acceleration 
at the beginning of the computation in the two opposite situation. 
In order to avoid the collision, whenever the car reaches line $S$ and selects an acceleration 
$A_c$, it has to check that the current state satisfies $\psi_{\textrm{sf}}'$ and does not satisfy 
$\psi_{\textrm{un}}'$. 
Figure~\ref{fig:analysis} depicts the evaluation of such formul\ae\ in a portion 
of the state space. 


\section{Conclusions}\label{conclusions}
In this paper we considered Michael's form hybrid automata, a class of automata particularly suitable
for approximations. On the basis of the observation that infinite precision of the models does not 
reflect real systems behaviors, we introduced and discussed different approximation techniques
over this class of automata. On the one hand, our comparison points out that disturbed automata cannot be 
formulated in terms of an $\epsilon$-semantics. As a matter of fact, $\epsilon$-semantics never over-approximate
standard semantics. On the other hand, we demonstrate that Fr\"anzle's approach can be modeled through a new 
semantics (tilde semantics) which provides an over-approximation of the original reach space of the hybrid automaton. 
However, it is important to notice that Fr\"anzle's reachability algorithm cannot be mapped into a completely symbolic 
algorithm (similar to the one presented in \cite{DEDS09}) since at each iteration it over-approximates the reached set, while
a symbolic algorithm would construct a new formula (nesting quantifiers) at each step and evaluate its 
approximated semantics only at the end of the computation. Hence, since quantified variables are never approximated, such a symbolic algorithm would over-approximate only  
the last step. 

Drawing inspiration from both disturbed hybrid automata and symbolic algorithms, we formalized a new 
$\epsilon$-semantics (bottom semantics) which plays a symmetrical role with respect to the introduction 
of noise.
If the disturbance of the continuous components expands trajectories, the application
of bottom semantics reduces it, under-approximating the reachability set. So,
we can say the bottom semantics describes a process of noise filtering in hybrid automata. 
In particular, since we exploit on bottom 
semantics the symbolic algorithmic approach described in \cite{DEDS09}, we use the same level of 
approximation for both halting conditions and output. Of course, reachability could be under-approximated 
using standard semantics and simply halting computation after a finite number of discrete steps. The 
meaning of our under-approximation is that we interpret bottom semantics as the ``correct'' semantics 
for noise filtering.

As future work we plan to extend our comparisons to other general frameworks for approximation 
techniques, such as the ones based on topology (see, e.g., \cite{Col05,Davoren}). In those frameworks 
instead of using distances among points, as we did, the authors defined distances among trajectories. 
Some basic differences between our approach and the one used in \cite{Col05} can be noticed 
considering the bouncing ball example presented both in \cite{DEDS09} and in \cite{Col05}. While in 
\cite{DEDS09} the infinite sequence of bounces cannot be observed since at a certain point these are 
smaller than the $\epsilon$-precision, in \cite{Col05} a compactification of the space is introduced to 
ensure convergence.

\bibliographystyle{eptcs}
\bibliography{biblio,hybridbib}

\begin{thebibliography}{10}
\providecommand{\bibitemdeclare}[2]{}
\providecommand{\surnamestart}{}
\providecommand{\surnameend}{}
\providecommand{\urlprefix}{Available at }
\providecommand{\url}[1]{\texttt{#1}}
\providecommand{\href}[2]{\texttt{#2}}
\providecommand{\urlalt}[2]{\href{#1}{#2}}
\providecommand{\doi}[1]{doi:\urlalt{http://dx.doi.org/#1}{#1}}
\providecommand{\bibinfo}[2]{#2}

\bibitemdeclare{inproceedings}{Alur}
\bibitem{Alur}
\bibinfo{author}{R.~\surnamestart Alur\surnameend},
  \bibinfo{author}{C.~\surnamestart Courcoubetis\surnameend},
  \bibinfo{author}{T.~A. \surnamestart Henzinger\surnameend} \&
  \bibinfo{author}{P.~H. \surnamestart Ho\surnameend} (\bibinfo{year}{1993}):
  \emph{\bibinfo{title}{Hybrid {A}utomata: {A}n {A}lgorithmic {A}pproach to the
  {S}pecification and {V}erification of {H}ybrid {S}ystems}}.
\newblock In: {\sl \bibinfo{booktitle}{Hybrid Systems}}, {\sl
  \bibinfo{series}{LNCS}} \bibinfo{volume}{736}, \bibinfo{publisher}{Springer},
  pp. \bibinfo{pages}{209--229}, \doi{10.1007/3-540-57318-6\_30}.

\bibitemdeclare{book}{bookselection}
\bibitem{bookselection}
\bibinfo{author}{J.~P. \surnamestart Aubin\surnameend} \&
  \bibinfo{author}{A.~\surnamestart Cellina\surnameend} (\bibinfo{year}{1984}):
  \emph{\bibinfo{title}{Differential Inclusions}}.
\newblock {\sl \bibinfo{series}{A Series of Comprehensive Studies in
  Mathematics}} \bibinfo{volume}{264}, \bibinfo{publisher}{Springer},
  \doi{10.1007/978-3-642-69512-4}.

\bibitemdeclare{inproceedings}{basu97}
\bibitem{basu97}
\bibinfo{author}{S.~\surnamestart Basu\surnameend} (\bibinfo{year}{1997}):
  \emph{\bibinfo{title}{{An Improved Algorithm for Quantifier Elimination Over
  Real Closed Fields}}}.
\newblock In: {\sl \bibinfo{booktitle}{{IEEE} Symposium on Foundations of
  Computer Science (FOCS'97)}}, \bibinfo{publisher}{IEEE Computer Society
  Press}, pp. \bibinfo{pages}{56--65}, \doi{10.1109/SFCS.1997.646093}.

\bibitemdeclare{inproceedings}{hsb2012}
\bibitem{hsb2012}
\bibinfo{author}{A.~\surnamestart Casagrande\surnameend},
  \bibinfo{author}{T.~\surnamestart Dreossi\surnameend} \&
  \bibinfo{author}{C.~\surnamestart Piazza\surnameend} (\bibinfo{year}{2012}):
  \emph{\bibinfo{title}{Hybrid Automata and $\epsilon$-Analysis on a Neural
  Oscillator}}.
\newblock In: {\sl \bibinfo{booktitle}{Proc. of the 1st International Workshop
  on Hybrid Systems and Biology}}, {\sl
  \bibinfo{series}{EPTCS}}~\bibinfo{volume}{92}, pp. \bibinfo{pages}{58--72},
  \doi{10.4204/EPTCS.92.5}.

\bibitemdeclare{article}{DEDS09}
\bibitem{DEDS09}
\bibinfo{author}{A.~\surnamestart Casagrande\surnameend},
  \bibinfo{author}{C.~\surnamestart Piazza\surnameend} \&
  \bibinfo{author}{A.~\surnamestart Policriti\surnameend}
  (\bibinfo{year}{2009}): \emph{\bibinfo{title}{{Discrete Semantics for Hybrid
  Automata}}}.
\newblock {\sl \bibinfo{journal}{Discrete Event Dynamic Systems}}
  \bibinfo{volume}{19}(\bibinfo{number}{4}), pp. \bibinfo{pages}{471--493},
  \doi{10.1007/s10626-009-0082-7}.

\bibitemdeclare{article}{focore2008}
\bibitem{focore2008}
\bibinfo{author}{A.~\surnamestart Casagrande\surnameend},
  \bibinfo{author}{C.~\surnamestart Piazza\surnameend},
  \bibinfo{author}{A.~\surnamestart Policriti\surnameend} \&
  \bibinfo{author}{B.~\surnamestart Mishra\surnameend} (\bibinfo{year}{2008}):
  \emph{\bibinfo{title}{Inclusion dynamics hybrid automata}}.
\newblock {\sl \bibinfo{journal}{Information and Computation}}
  \bibinfo{volume}{206}(\bibinfo{number}{12}), pp. \bibinfo{pages}{1394--1424},
  \doi{10.1016/j.ic.2008.09.001}.

\bibitemdeclare{techreport}{Col05}
\bibitem{Col05}
\bibinfo{author}{Pieter \surnamestart Collins\surnameend}
  (\bibinfo{year}{2005}): \emph{\bibinfo{title}{{Hybrid Trajectory Spaces}}}.
\newblock \bibinfo{type}{Technical Report}, \bibinfo{institution}{Centrum voor
  Wiskunde en Informatica (CWI)}.

\bibitemdeclare{unpublished}{Davoren}
\bibitem{Davoren}
\bibinfo{author}{J.~\surnamestart Davoren\surnameend} \&
  \bibinfo{author}{I.~\surnamestart Epstein\surnameend} (\bibinfo{year}{2008}):
  \emph{\bibinfo{title}{Topologies, Convergence, and Uniformities in General
  Hybrid Path Spaces,}}.
\newblock \bibinfo{note}{Preprint}.

\bibitemdeclare{inproceedings}{franzle99}
\bibitem{franzle99}
\bibinfo{author}{M.~\surnamestart Fr{\"a}nzle\surnameend}
  (\bibinfo{year}{1999}): \emph{\bibinfo{title}{{Analysis of Hybrid Systems: An
  ounce of realism can save an infinity of states}}}.
\newblock In: {\sl \bibinfo{booktitle}{Computer Science Logic (CSL'99)}}, {\sl
  \bibinfo{series}{LNCS}} \bibinfo{volume}{1683},
  \bibinfo{publisher}{Springer}, pp. \bibinfo{pages}{126--140},
  \doi{10.1007/3-540-48168-0\_10}.

\bibitemdeclare{article}{girard}
\bibitem{girard}
\bibinfo{author}{A.~\surnamestart Girard\surnameend}, \bibinfo{author}{A.~A.
  \surnamestart Julius\surnameend} \& \bibinfo{author}{G.~J. \surnamestart
  Pappas\surnameend} (\bibinfo{year}{2008}): \emph{\bibinfo{title}{{Approximate
  Simulation Relations for Hybrid Systems}}}.
\newblock {\sl \bibinfo{journal}{Discrete Event Dynamic Systems}}
  \bibinfo{volume}{18}(\bibinfo{number}{2}), pp. \bibinfo{pages}{163--179},
  \doi{10.1007/s10626-007-0029-9}.

\bibitemdeclare{inproceedings}{undecidable}
\bibitem{undecidable}
\bibinfo{author}{T.~A. \surnamestart Henzinger\surnameend},
  \bibinfo{author}{P.~W. \surnamestart Kopke\surnameend},
  \bibinfo{author}{A.~\surnamestart Puri\surnameend} \&
  \bibinfo{author}{P.~\surnamestart Varaiya\surnameend} (\bibinfo{year}{1995}):
  \emph{\bibinfo{title}{What's decidable about hybrid automata?}}
\newblock In: {\sl \bibinfo{booktitle}{Proc. of ACM Symposium on Theory of
  Computing (STOCS'95)}}, \bibinfo{publisher}{ACM}, pp.
  \bibinfo{pages}{373--382}, \doi{10.1145/225058.225162}.

\bibitemdeclare{inproceedings}{DBLP:conf/hybrid/HenzingerR00}
\bibitem{DBLP:conf/hybrid/HenzingerR00}
\bibinfo{author}{T.~A. \surnamestart Henzinger\surnameend} \&
  \bibinfo{author}{J.-F. \surnamestart Raskin\surnameend}
  (\bibinfo{year}{2000}): \emph{\bibinfo{title}{Robust Undecidability of Timed
  and Hybrid Systems}}.
\newblock In: {\sl \bibinfo{booktitle}{Proc. of the 3rd International Workshop
  Hybrid Systems: Computation and Control (HSCC'00)}}, {\sl
  \bibinfo{series}{LNCS}} \bibinfo{volume}{1790},
  \bibinfo{publisher}{Springer}, pp. \bibinfo{pages}{145--159},
  \doi{10.1007/3-540-46430-1\_15}.

\bibitemdeclare{article}{ominimal}
\bibitem{ominimal}
\bibinfo{author}{G.~\surnamestart Lafferriere\surnameend},
  \bibinfo{author}{G.~J. \surnamestart Pappas\surnameend} \&
  \bibinfo{author}{S.~\surnamestart Sastry\surnameend} (\bibinfo{year}{2000}):
  \emph{\bibinfo{title}{{O-minimal Hybrid Systems}}}.
\newblock {\sl \bibinfo{journal}{Mathematics of Control, Signals, and Systems}}
  \bibinfo{volume}{13}, pp. \bibinfo{pages}{1--21}, \doi{10.1007/PL00009858}.

\bibitemdeclare{article}{lafferiere01}
\bibitem{lafferiere01}
\bibinfo{author}{G.~\surnamestart Lafferriere\surnameend},
  \bibinfo{author}{G.~J. \surnamestart Pappas\surnameend} \&
  \bibinfo{author}{S.~\surnamestart Yovine\surnameend} (\bibinfo{year}{2001}):
  \emph{\bibinfo{title}{{Symbolic Reachability Computation for Families of
  Linear Vector Fields}}}.
\newblock {\sl \bibinfo{journal}{J. Symb. Comput.}}
  \bibinfo{volume}{32}(\bibinfo{number}{3}), pp. \bibinfo{pages}{231--253},
  \doi{10.1006/jsco.2001.0472}.

\bibitemdeclare{book}{topology}
\bibitem{topology}
\bibinfo{author}{B.~\surnamestart Mendelson\surnameend} (\bibinfo{year}{1990}):
  \emph{\bibinfo{title}{{Introduction to Topology}}}, \bibinfo{edition}{{III}}
  edition.
\newblock \bibinfo{publisher}{Dover Books on Mathematics}.

\bibitemdeclare{book}{mendel}
\bibitem{mendel}
\bibinfo{author}{E.~\surnamestart Mendelson\surnameend} (\bibinfo{year}{1997}):
  \emph{\bibinfo{title}{{Introduction to Mathematical Logic}}},
  \bibinfo{edition}{{IV}} edition.
\newblock \bibinfo{publisher}{CRC Press}.

\bibitemdeclare{techreport}{ourpaper}
\bibitem{ourpaper}
\bibinfo{author}{C.~\surnamestart Piazza\surnameend},
  \bibinfo{author}{M.~\surnamestart Antoniotti\surnameend},
  \bibinfo{author}{V.~\surnamestart Mysore\surnameend},
  \bibinfo{author}{A.~\surnamestart Policriti\surnameend},
  \bibinfo{author}{F.~\surnamestart Winkler\surnameend} \&
  \bibinfo{author}{B.~\surnamestart Mishra\surnameend} (\bibinfo{year}{2005}):
  \emph{\bibinfo{title}{{Algorithmic Algebraic Model Checking I: The Case of
  Biochemical Systems and their Reachability Analysis}}}.
\newblock \bibinfo{type}{CIMS-TR} \bibinfo{number}{2005-859},
  \bibinfo{institution}{Courant Institute Of Mathematical Sciences}.

\bibitemdeclare{inproceedings}{DBLP:conf/rtss/PrabhakarVVD09}
\bibitem{DBLP:conf/rtss/PrabhakarVVD09}
\bibinfo{author}{P.~\surnamestart Prabhakar\surnameend},
  \bibinfo{author}{V.~\surnamestart Vladimerou\surnameend},
  \bibinfo{author}{M.~\surnamestart Viswanathan\surnameend} \&
  \bibinfo{author}{G.~E. \surnamestart Dullerud\surnameend}
  (\bibinfo{year}{2009}): \emph{\bibinfo{title}{Verifying Tolerant Systems
  Using Polynomial Approximations}}.
\newblock In: {\sl \bibinfo{booktitle}{Proc. of the 30th IEEE Real-Time Systems
  Symposium (RTSS'09)}}, \bibinfo{publisher}{IEEE Computer Society Press}, pp.
  \bibinfo{pages}{181--190}, \doi{10.1109/RTSS.2009.28}.

\bibitemdeclare{incollection}{ratschan09}
\bibitem{ratschan09}
\bibinfo{author}{S.~\surnamestart Ratschan\surnameend} (\bibinfo{year}{2010}):
  \emph{\bibinfo{title}{Safety Verification of Non-linear Hybrid Systems Is
  Quasi-Semidecidable}}.
\newblock In: {\sl \bibinfo{booktitle}{Proc. of the 7th Conference on Theory
  and Applications of Models of Computation (TAMC'10)}}, {\sl
  \bibinfo{series}{LNCS}} \bibinfo{volume}{6108},
  \bibinfo{publisher}{Springer}, pp. \bibinfo{pages}{397--408},
  \doi{10.1007/978-3-642-13562-0\_36}.

\bibitemdeclare{inproceedings}{Sturm:2011:VSU:1993886.1993935}
\bibitem{Sturm:2011:VSU:1993886.1993935}
\bibinfo{author}{T.~\surnamestart Sturm\surnameend} \&
  \bibinfo{author}{A.~\surnamestart Tiwari\surnameend} (\bibinfo{year}{2011}):
  \emph{\bibinfo{title}{Verification and synthesis using real quantifier
  elimination}}.
\newblock In: {\sl \bibinfo{booktitle}{Proc. of the 36th international
  symposium on Symbolic and algebraic computation (ISSAC'11)}},
  \bibinfo{publisher}{ACM}, pp. \bibinfo{pages}{329--336},
  \doi{10.1145/1993886.1993935}.

\bibitemdeclare{book}{tarski}
\bibitem{tarski}
\bibinfo{author}{A.~\surnamestart Tarski\surnameend} (\bibinfo{year}{1951}):
  \emph{\bibinfo{title}{A {D}ecision {M}ethod for {E}lementary {A}lgebra and
  {G}eometry}}.
\newblock \bibinfo{publisher}{Univ. California Press}.

\bibitemdeclare{inproceedings}{tiwari}
\bibitem{tiwari}
\bibinfo{author}{A.~\surnamestart Tiwari\surnameend} \&
  \bibinfo{author}{G.~\surnamestart Khanna\surnameend} (\bibinfo{year}{2002}):
  \emph{\bibinfo{title}{Series of {A}bstractions for {H}ybrid {A}utomata}}.
\newblock In: {\sl \bibinfo{booktitle}{Proc. of Hybrid Systems: Computation and
  Control (HSCC'02)}}, {\sl \bibinfo{series}{LNCS}} \bibinfo{volume}{2289},
  \bibinfo{publisher}{Springer}, pp. \bibinfo{pages}{465--478},
  \doi{10.1007/3-540-45873-5\_36}.

\end{thebibliography}

\end{document}